\documentclass[master]{ws-rv9x6} 
\usepackage{ws-rv-har}    
\usepackage{ws-rv-thm}    
\usepackage{subfigure}    
\usepackage{ws-index}     
\usepackage{enumitem}
\usepackage[percent]{overpic}
\usepackage{xcolor}
\makeindex
\newindex{aindx}{adx}{and}{Author Index}       
\renewindex{default}{idx}{ind}{Subject Index}  

\begin{document}		
\newcommand{\ltsima}{$\; \buildrel < \over \sim \;$}
\newcommand{\lsim}{\lower.5ex\hbox{\ltsima}}
\newcommand{\gtsima}{$\; \buildrel > \over \sim \;$}
\newcommand{\gsim}{\lower.5ex\hbox{\gtsima}}
\newcommand{\bra}{\langle}
\newcommand{\ket}{\rangle}
\newcommand{\lprime}{\ell^\prime}
\newcommand{\lpp}{\ell^{\prime\prime}}
\newcommand{\mprime}{m^\prime}
\newcommand{\mpp}{m^{\prime\prime}}
\newcommand{\ci}{\mathrm{i}}
\newcommand{\dd}{\mathrm{d}}
\newcommand{\veck}{\mathbf{k}}
\newcommand{\vecx}{\mathbf{x}}
\newcommand{\vecr}{\mathbf{r}}
\newcommand{\vecv}{\mathbf{\upsilon}}
\newcommand{\vecw}{\mathbf{\omega}}
\newcommand{\vecj}{\mathbf{j}}
\newcommand{\vecq}{\mathbf{q}}
\newcommand{\vecl}{\mathbf{l}}
\newcommand{\vecn}{\mathbf{n}}
\newcommand{\lm}{\ell m}
\newcommand{\that}{\hat{\theta}}
\newcommand{\thatp}{\that^\prime}
\newcommand{\chip}{\chi^\prime}
\newcommand{\hs}{\hspace{1mm}}
\newcommand{\nar}{New Astronomy Reviews}
\def\gsim{~\rlap{$>$}{\lower 1.0ex\hbox{$\sim$}}}
\def\lsim{~\rlap{$<$}{\lower 1.0ex\hbox{$\sim$}}}
\def\Msun {\,\mathrm{M}_\odot}
\def\Jcrit {J_\mathrm{crit}}
\newcommand{\rsun}{R_{\odot}}
\newcommand{\mbh}{M_{\rm BH}}
\newcommand{\Msunyr}{M_\odot~{\rm yr}^{-1}}
\newcommand{\mdot}{\dot{M}_*}
\newcommand{\ledd}{L_{\rm Edd}}
\newcommand{\cmc}{{\rm cm}^{-3}}
\def\gsim{~\rlap{$>$}{\lower 1.0ex\hbox{$\sim$}}}
\def\lsim{~\rlap{$<$}{\lower 1.0ex\hbox{$\sim$}}}
\def\Msun {\,\mathrm{M}_\odot}
\def\Jcrit {J_\mathrm{crit}}

\def\simgreat{\lower2pt\hbox{$\buildrel {\scriptstyle >}
   \over {\scriptstyle\sim}$}}
\def\simless{\lower2pt\hbox{$\buildrel {\scriptstyle <}
   \over {\scriptstyle\sim}$}}
\def\msobh{M_\bullet^{\rm sBH}}
\def\zodot{\,{\rm Z}_\odot}
\newcommand{\lambdabar}{\mbox{\makebox[-0.5ex][l]{$\lambda$} \raisebox{0.7ex}[0pt][0pt]{--}}}

\def\na{NewA}%
\def\aj{AJ}%
\def\araa{ARA\&A}%
\def\apj{ApJ}%
\def\apjl{ApJ}%
\def\jcap{JCAP}

\def\pasa{PASA}

\def\apjs{ApJS}%
\def\ao{Appl.~Opt.}%
\def\apss{Ap\&SS}%
\def\aap{A\&A}%
\def\aapr{A\&A~Rev.}%
\def\aaps{A\&AS}%
\def\azh{AZh}%
\def\baas{BAAS}%
\def\jrasc{JRASC}%
\def\memras{MmRAS}%
\def\mnras{MNRAS}%
\def\pra{Phys.~Rev.~A}%
\def\prb{Phys.~Rev.~B}%
\def\prc{Phys.~Rev.~C}%
\def\prd{Phys.~Rev.~D}%
\def\pre{Phys.~Rev.~E}%
\def\prl{Phys.~Rev.~Lett.}%
\def\pasp{PASP}%
\def\pasj{PASJ}%
\def\qjras{QJRAS}%
\def\skytel{S\&T}%
\def\solphys{Sol.~Phys.}%

\def\sovast{Soviet~Ast.}%
\def\ssr{Space~Sci.~Rev.}%
\def\zap{ZAp}%
\def\nat{Nature}%
\def\iaucirc{IAU~Circ.}%
\def\aplett{Astrophys.~Lett.}%
\def\apspr{Astrophys.~Space~Phys.~Res.}%
\def\bain{Bull.~Astron.~Inst.~Netherlands}%
\def\fcp{Fund.~Cosmic~Phys.}%
\def\gca{Geochim.~Cosmochim.~Acta}%
\def\grl{Geophys.~Res.~Lett.}%
\def\jcp{J.~Chem.~Phys.}%
\def\jgr{J.~Geophys.~Res.}%
\def\jqsrt{J.~Quant.~Spec.~Radiat.~Transf.}%
\def\memsai{Mem.~Soc.~Astron.~Italiana}%
\def\nphysa{Nucl.~Phys.~A}%

\def\physrep{Phys.~Rep.}%
\def\physscr{Phys.~Scr}%
\def\planss{Planet.~Space~Sci.}%
\def\procspie{Proc.~SPIE}%

\newcommand{\rmp}{Rev. Mod. Phys.}
\newcommand{\ijmpd}{Int. J. Mod. Phys. D}
\newcommand{\sovjetp}{Soviet J. Exp. Theor. Phys.}
\newcommand{\jkas}{J. Korean. Ast. Soc.}
\newcommand{\PPVI}{Protostars and Planets VI}
\newcommand{\njp}{New J. Phys.}
\newcommand{\rap}{Res. Astro. Astrophys.}

\title{Formation of the First Black Holes}

\setcounter{chapter}{11}

\chapter{Formation of the First Black Holes:\newline Current observational status$^1$}

\author[Dominik R.G. Schleicher]{Dominik R.G. Schleicher}

\address{Astronomy Department, \\Universidad de Concepci\'on, \\Barrio Universitario, \\ Concepci\'on, Chile, \\ dschleicher@astro-udec.cl}

\begin{abstract}
In this chapter, we review the current observational status of the first supermassive black holes. It is clear that such a review can hardly be complete, due to the wealth of surveys that has been pursued, including different wavelengths and different observational techniques. This chapter will focus on the main results that have been obtained,  considering the detections of $z\gtrsim6$ supermassive black holes in large surveys such as SDSS, CFHQS and Pan-STARRS. In addition, we will discuss upper limits and constraints on the population of the first black holes that can be derived from observational data, in particular in the X-ray regime, as these provide additional relevant information for the comparison with formation scenarios. 
\end{abstract}

\body

\setcounter{page}{223}

\section{Introduction}

Having discussed the\footnotetext{$^1$ Preprint~of~a~review volume chapter to be published in Latif, M., \& Schleicher, D.R.G., ''Current Observational Status'', Formation of the First Black Holes, 2018 \textcopyright Copyright World Scientific Publishing Company, https://www.worldscientific.com/worldscibooks/10.1142/10652 } formation of the first black holes, their initial mass function as well as the subsequent growth in previous chapters, it is now time to review the current observational status regarding the first supermassive black holes. The main focus here is on the black holes themselves, not on their host galaxies, which have been discussed in detail in other reviews such as \citet{Gallerani17}. In the following, we will start presenting the main surveys that have identified quasars at $z\gtrsim6$ in section~\ref{surveys}. Upper limits and constraints from x-ray surveys and stacking techniques are presented in section~\ref{xray}, along with the constraints derived from them on the population of the first black holes.

\section{Surveys of $z\gtrsim6$ supermassive black holes}\label{surveys}
The currently known quasars at $z\gtrsim6$ correspond to the most direct information that is available on the high-redshift black hole population, providing constraints on formation and accretion models through the masses of the supermassive black holes as well as the redshift at which they exist. In the following, we will provide an overview of the main surveys that have contributed to their detection.

\subsection{The Sloan Digital Sky Survey}
The Sloan Digital Sky Survey (SDSS)\footnote{Webpage SDSS: http://www.sdss.org/} \cite{York00} is one of the largest and longest astronomical surveys running so far, since the year 2000. In the initial configuration over the first five years (SDSS-I), it carried out deep multi-color imaging in the u, g, r, i and z-band over $8000$ square degrees and measured spectra of more than $700,000$ astronomical objects (see \citet{Fukugita96} for the description of the photometric system). Within the second period from 2005-2008 (SDSS-II), it completed the original survey goals to imagine half of the northern sky and to map the 3-dimensional clustering of one million galaxies and $100,000$ quasars. In 2008-2014 (SDSS-III), after a major upgrade of the spectrographs and with two new instruments to execute a set of four surveys, it mapped the clustering of galaxies and intergalactic gas in the distant Universe through the BOSS survey. The current generation of the SDSS (SDSS-IV, 2014-2020) includes many programs, most relevantly here the SPectroscopic IDentification of EROSITA Sources (SPIDERS) that will provide an unique census of supermassive black-hole and large scale structure growth, targeting X-ray sources from ROSAT, XMM and eROSITA.

\begin{figure}
\includegraphics[scale=0.5]{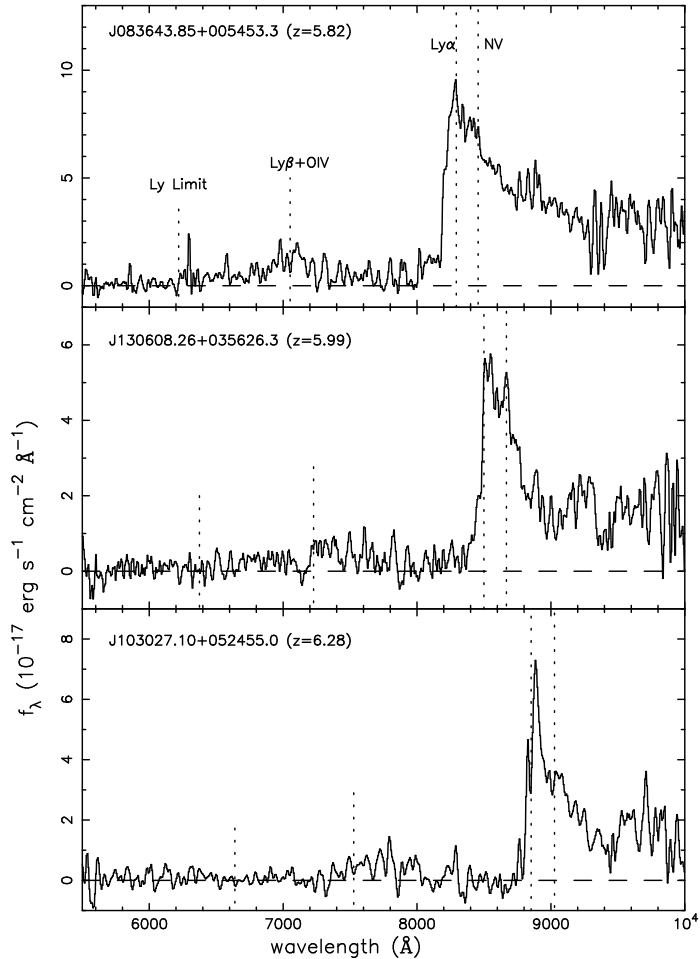}
\caption{The spectra of the three new $z>5.8$ SDSS quasars discovered by \citet{Fan01}. Figure adopted from \citet{Fan01}, \textcopyright AAS. Reproduced with permission.}
\label{sdss}
\end{figure}

Overall and especially in the beginning, SDSS has played a central role in identifying quasars at $z\gtrsim6$. Many of them were discovered from the $\sim8000$~deg$^2$ imaging data, where luminous quasars ($z_{AB}\leq 20$) were selected as i-dropout objects through optical colors. Near-infrared (NIR) photometry as well as optical spectroscopy was subsequently used to distinguish them from late-type dwarfs \citep{Fan01}. A first $z=5.80$ quasar, SDSSp J104433.04-012502.2, was discovered in 2000 by SDSS \citet{Fan00}, and many more were found in subsequent years. Three additional objects, SDSSp J083643.85+005453.3 ($z=5.82$), J130608.26+035626.3 ($z=5.99$), and J103027.10+052455.0 ($z=6.28$) were discovered in 2001 \citep[see also Fig.~\ref{sdss}][]{Fan01}. These quasars were subsequently used to put constraints onto the end of reionization, and particularly the existence of a complete  Lyman~$\alpha$ Gunn-Peterson (GP) trough in the spectrum of the $z=6.28$ quasar allowed to place strong limits \citep{Fan02}. 

Three additional quasars, all at $z>6$, have been discovered in 2003:  J114816.64+525150.3 ($z=6.43$), J104845.05+463718.3 ($z=6.23$), and J163033.90+401209.6 ($z=6.05$) \citep{Fan2003}. These objects correspond to the bright-end slope of the quasar luminosity function, with $M_{1450}<-26.8$ (magnitude at $1450$~\AA). Their co-moving number density was thus estimated to be $(8\pm3)\times10^{-10}$~Mpc$^{-3}$. The discovery of 5 additional objects at $z>5.7$ has led to improved comoving number density estimates for objects with $M_{1450}<-26.7$ of $(6\pm2)\times10^{-10}$~Mpc$^{-3}$ at $z\sim6$ \citep{Fan04}. $7$ quasars were subsequently discovered at $z>5.7$, with redshifts between $5.79$ and $6.13$. Two of them were at $z>6$ \citep{Fan2006}. Based on such detections, a sample of $19$ quasars at $z\sim6$ could be established to probe reionization in 2006 \citep{Fan06b}.

With the discovery of $5$ additional quasars at $z\sim6$ selected from a $260$~deg$^2$ field (one of them also independently discovered by the UKIRT Infrared Deep Sky Survey)\footnote{Webpage UKIDSS: http://www.ukidss.org/}, a  complete flux-limited quasar sample at $z_{AB} < 21$ could be established, consisting of a total of 12 objects \citep{Jiang08}. This sample spans a redshift range of $5.85\leq z \leq 6.12$ and a luminosity range of  $-26.5 \leq M_{1450} \leq -25.4$. The co-moving number density per luminosity magnitude was estimated to be $(5.0 \pm 2.1) \times 10^{-9}$~Mpc$^{-3}$~mag$^{-1}$. The sample allowed the first assessment of the bright-end quasar luminosity function at $z\sim6$ as a power law, $\Phi(L_{1450})\propto L_{1450}^\beta$, with $\beta=-3.1\pm0.4$. The subsequent detection of six additional quasars at $z\sim6$ at at $21 < z_{AB} < 21.8$ essentially confirmed these conclusions \citep{Jiang09}. 

The final sample based on SDSS discoveries alone consisted of $52$ quasars with $5.7\leq z\leq 6.4$,  spanning a wide luminosity range of$-29.0\leq {M}_{1450}\leq -24.5$ \citep{Jiang16}. The bright-end-slope is well constrained to be $\beta=-2.8\pm0.2$. Parametrizing the spatial density of luminous quasars as $\rho(M_{1450})=\rho(z=6)\times10^{k(z-6)}$, they found that it drops rapidly from $z\sim5$ to $z\sim6$ with $k=-0.72\pm0.11$. Due to the small number density, it was shown that the observed population of quasars cannot provide enough photons to account for cosmic reionization, even though of course contributions from lower-luminosity quasars may be present. Overall, SDSS has opened up an important window to study the population of supermassive black holes at $z\sim6$, including important objects such as the quasars at $z=6.28$ and $z=6.40$, which were included in many follow-up investigations. A particularly relevant recent discovery is the quasar SDSS J010013.02+280225.8 at $z=6.30$ with an estimated black hole mass of $1.2\times10^{10}$~M$_\odot$ \citep{Wu15}, and thus the most massive supermassive black holes at these redshifts so far.

\subsection{The Canada-France high-z quasar survey}
The Canada-France high-z quasar survey (CFHQS) makes use of optical imaging pursued in the Canada-France-Hawaii Telescope (CFHT) Legacy Survey\footnote{Webpage: http://www.cfht.hawaii.edu/Science/CFHTLS/}. It uses several different data sets for the search for high-redshift quasars: The  majority  of  the  sky  area  ($\sim550$~deg$^2$) is   part   of   the   RCS-2   survey  \citep{Yee07}, and  consists  of  MegaCam g$^\prime$, r$^\prime$, i$^\prime$, z$^\prime$ imaging data with  exposure times  in  each  filter  of  240~s,  480~s,  500~s,  360~s,  respectively. The  CFHT  Legacy  Survey  (CFHTLS)  Very  Wide covers  several hundred square degrees, of which $\sim150$~deg$^2$ are included in the CFHQS. The total exposure times for the Very Wide are comparable to RCS-2 with 540~s at i$^\prime$ and 420~s at z$^\prime$. The CFHTLS Wide is the intermediate depth per area component of the CFHTLS. It consists of $171$~deg$^2$ at u$^\prime$, g$^\prime$, r$^\prime$, i$^\prime$, z$^\prime$ with typical MegaCam integration times of $4300$~s at i$^\prime$ and $3600$~s at z$^\prime$. The CFHTLS Deep consists of four MegaCam pointings each of $\sim1$~deg$^2$ at u$^\prime$ g$^\prime$ r$^\prime$ i$^\prime$ z$^\prime$, with typical integration times of $250.000$~s at i$^\prime$ and $200.000$~s at z$^\prime$. Finally, the Subaru/XMM-Newton Deep Survey (SXDS) is a deep B, V, R, i$^\prime$, z$^\prime$ survey of $\sim1.2$~deg$^2$ carried out at the Subaru 8.2m optical/infrared Telescope\footnote{Webpage Subaru: https://subarutelescope.org/} \citep{Furusawa08}. SXDS employed typical integration times of $25.000$~s at $i^\prime$ and $13.000$~s at $z^\prime$ and is thus comparable to the CFHTLS Deep.

The first results of the survey yielded 24 quasar candidates at redshifts $5.7<z<6.4$ \citep{Willott05}, thereby yielding constraints and upper limits on the  quasar luminosity function. In 2007, the survey discovered four quasars above redshift $6$, including the highest-redshift quasar at that time, CFHQS J2329-0301 at $z=6.43$ \citep{Willott2007}. Six additional quasars above $z\geq5.9$ were subsequently found, with luminosities $10-75$ times lower than the most luminous Sloan Digital Sky Survey quasars at this redshift \citep{Willott09}. The least luminous among them, CFHQS J0216-0455 at z = 6.01 with an absolute magnitude $M_{1450} = -22.21$ is well below the expected break in the luminosity function where a transition from the steep power-law is expected to occur. Finally, nine additional $z\sim6$ quasars were found in 2010, thus bringing the total number of CFHQS quasars to $19$ \citep{Willot2010}. Their binned luminosity function suggest a break  at $M_{1450}\sim-25$, with a double power-law maximum likelihood fit to the data being consistent with the binned results. The CFHQS was thus particularly important to complement SDSS results through the detection of quasars at lower luminosity. 

\begin{figure*}
\includegraphics[scale=0.4]{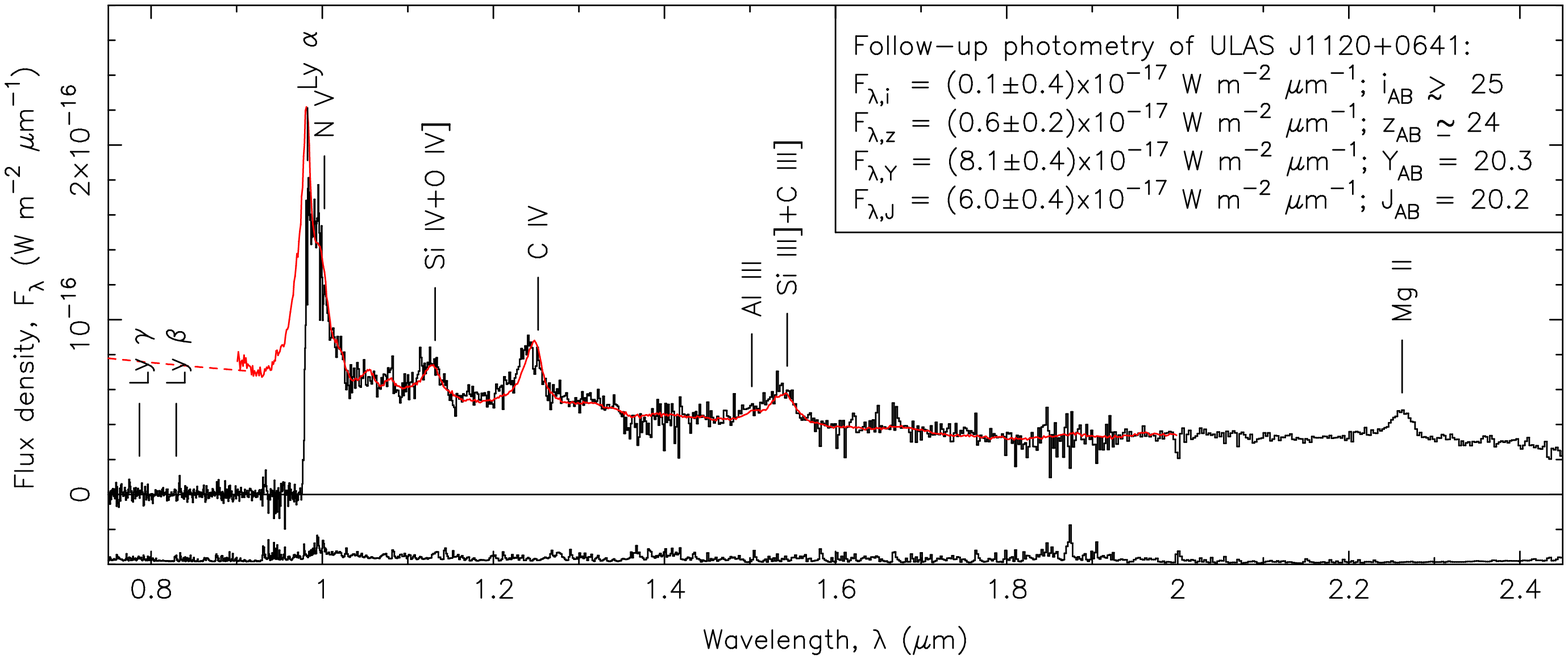}
\caption{Spectrum of the $z=7.085$ quasar discovered by \citet{Mortlock2011}. Figure adopted from \citet{Mortlock2011}. }
\label{spectrum}
\end{figure*}

\subsection{The UKIRT Infrared Deep Sky Survey}
While optical surveys were extremely important to find high-redshift quasars at $\sim6$, the same surveys were insensitive to sources beyond $z=6.5$, providing a strong limitation to find earlier objects. For the search of higher-redshift objects, near-infrared surveys are thus very important. The UKIRT Infrared Deep Sky Survey (UKIDSS)\footnote{Webpage UKIDSS: http://www.ukidss.org/} was the first near-infrared survey with the capability to find quasars at $z>6.5$. It is the successor to the 2MASS\footnote{Webpage 2MASS: https://www.ipac.caltech.edu/2mass/}, and began in 2005. Its goal is to survey $7500$~deg$^2$ of the Northern sky, extending over both high and low Galactic latitudes. The depth is three magnitudes deeper than 2MASS (K=18.3), thus forming the near-infrared counterpart to the Sloan Digital Sky Survey. The survey instrument is WFCAM on the UK Infrared Telescope (UKIRT)\footnote{Webpage UKRIT: http://www.ukirt.hawaii.edu/} in Hawaii. One of the science goals of the survey is finding the highest-redshift quasars at $z\sim7$ \citep{Lawrence07}.

A first exploratory search for quasars at $z\sim6-8$ has been performed in 2007 in the Early Data Release, initially yielding 34 candidates, which were however shown to be brown dwarfs via spectroscopic follow-up \citep{Glikman08}. The first discovery of a luminous $z\sim6$ quasar (ULAS J020332.38+001229.2) from near-infrared data was reported by \citet{Venemans07}, at a redshift of $z=5.86$.  Subsequently, the discovery of the $z=6.13$ quasar ULAS J131911.29+095051.4 was announced by \citet{Mortlock09}. With ULAS J112001.48+064124.3, an even higher-redshift quasar, and for about 7 years the highest redshift quasar, was subsequently discovered at $z=7.085$, i.e. about 770~million years after the Big Bang \citep{Mortlock2011}. The quasar has a high luminosity of $6.3\times10^{13}$~$L_\odot$, corresponding to a black hole mass of about $2\times10^9$~M$_\odot$. The presence of such an object emphasizes the very limited time that is needed to grow such massive black holes, though requiring an average accretion rate of the order $2.6$~M$_\odot$~yr$^{-1}$. The spectra shown in Fig.~\ref{spectrum} exhibit strong metal lines similar to the brightest quasars at lower redshift, and indicate that metal enrichment in its central region has been very efficient irrespective of the high redshift. Only recently, an even higher redshift supermassive black hole has been discovered, with a mass of $\sim8\times10^8$~M$_\odot$ at $z=7.54$ \citep{Banados18}. The redshift of the source corresponds to a time of 700 million years after the Big Bang. This discovery was based on data from UKIDSS, combined with the AllWISE survey\footnote{Website AllWISE: http://wise2.ipac.caltech.edu/docs/release/allwise/} (a successor of the WISE survey\footnote{Webpage WISE: https://www.nasa.gov/mission$\_$pages/WISE/main/index.html}), as well as DECam data (see subsection~\ref{DECam} for more information on DECam).

\begin{figure*}[ht]
\centering
\includegraphics[scale=0.50]{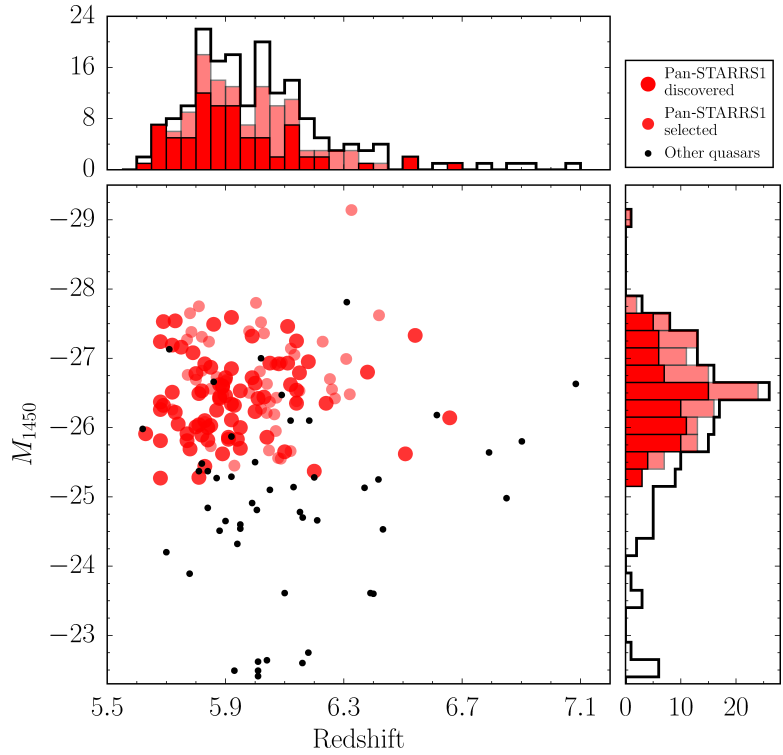}
\caption{
Redshift and absolute UV magnitude ($M_{1450}$) distribution of all the known $z>5.6$ quasars as of 2016 March \citep{Banados16}. Figure adopted from \citet{Banados16},  \textcopyright AAS. Reproduced with permission.
}
\label{fig:banados}
 \end{figure*}

\subsection{The Pan-STARRS distant $z>5.6$ quasar survey}
The Panoramic Survey Telescope and Rapid Response System (Pan-STARRS)\footnote{Webpage Pan-STARRS: https://panstarrs.stsci.edu/} is a system for wide-field astronomical imaging at the Institute for Astronomy of the University of Hawaii. Pan-STARRS1 (PS1) was the first part of Pan-STARRS to be completed. It used a $1.8$~meter telescope along with a $1.4$~Gigapixel camera  to image the sky in five broadband  filters (g, r, i, z, y), with $5\sigma$ limiting magnitudes of $(23.2, 23.0, 22.7, 22.1, 21.1)$. While PS1 is now completed, the focus has shifted towards PS2, where first light was achieved in 2013. The overall design foresees  up to 4 possible telescopes.

The PS1 survey \citep{Kaiser02, Kaiser10} has imaged the whole sky above a declination of $-30^\circ$ for about four years, which has led to a wealth of data and a large amount of new discoveries. The first high-redshift quasar discovered through the survey was announced in 2012, with a redshift of $z=5.73$ and an estimated black hole mass of $6.9\times10^9$~M$_\odot$ \citep{Morganson12}. \citet{Banados2014} subsequently discovered $8$ new quasars at $5.7 \leq z \leq 6.0$ with a range of luminosities, $-27.3 \leq M_{1450} \leq -25.4$, increasing the number of known quasars at $z>5.7$ by $10\%$ at that time. \citet{Venemans2015} subsequently announced 3 new quasars at $z=6.50$, $6.52$, and $6.66$, thereby substantially adding to the number of $z>6.5$ objects, including the brightest such object reported so far with $M_{1450}=-27.4$, the quasar PSO J036.5078 + 03.0498.

With improved selection criteria, \citet{Banados16} identified $63$ additional quasars at $5.6\leq z\leq 6.7$, leading to a total of 77 quasars that were identified through PS1. The total PS1 sample, which includes quasars previously discovered through other surveys, comprises a total of $124$~quasars and spans a factor of $\sim20$ in quasar luminosity. As such, it has considerably extended the number of known quasars at such redshift. The redshift and UV magnitude distribution of the currently known sample is given in Fig.~\ref{fig:banados}. {Recently, \citet{Mazzu17} discovered 6 additional $z\gtrsim6.5$ quasars via Pan-STARRS, and obtained a sample of 15 $z\gtrsim 6.5$ quasars to perform a homogeneous and comprehensive analysis. They derived typical black hole masses of $(0.3-5)\times10^9$~M$_\odot$ and Eddington  ratios of at least $0.39$ or higher. In addition,  the majority of $z\gtrsim 6.5$ quasars show large blueshifts of the broad C IV $\lambda$ 1549 emission line compared to the systemic redshift of the quasars, with a median value $\sim3\times$ higher than a quasar sample at $z\sim1$.
}

\subsection{The Subaru High-z Exploration of Low-Luminosity Quasars survey}
The Subaru High-z Exploration of Low-Luminosity Quasars (SHELLQs) survey is the firsts $1000$~deg$^2$ class survey for high-z quasars with a $8$~m class telescope like Subaru\footnote{Webpage Subaru: https://subarutelescope.org/}. The survey exploits multiband photometry data produced by the Subaru Hyper Suprime-Cam (HSC) \citep{Miyazaki12}, as part of the Subaru Strategic Program (SPP) survey. The HSC-SPP survey is a large collaborative project including researchers from Japan, Taiwan and Princeton. It started in early 2014, and will include 300 observing nights, lasting until 2019. It uses the HSC, a wide-field camera equipped with 116 2K $\times$ 4K Hamamatsu fully depleted CCDs, of  which 104 are used to obtain science data. Within the HSC-SPP, the Wide layer aims to observe $1400$~deg$^2$ mostly along the equator through five broad-band filters, aiming to reach a $5\sigma$ limiting magnitude of $g=26.5$, $r=26.1$, $i=25.9$, $z=25.1$, $y=24.4$.~mag within $2.''0$ apertures. The Deep and Ultra-Deep layers observed 27 and 3.5~deg$^2$, using 5 broad-band and 4 narrow-band filters, aiming to reach a $5\sigma$ limiting depth of $r=27.1$~mag (Deep) or $27.7$~mag (Ultra-Deep). 

SHELLQs exploits the HSC data to search for low-luminosity quasars at high redshift. The filter set is sensitive to quasars with redshifts up to $z\sim7.4$, i.e. beyond the current quasar redshift record, even though the detection capability sharply drops at $z>7$ as the Gunn-Peterson effect, so that the survey is limited to very luminous objects at those redshifts. The survey discovered $15$ quasars and bright galaxies at redshifts $5.7<z<6.9$, of which six were determined to be likely quasars \citep{Matsuoka16}. Subsequently the survey presented the spectroscopic identification of $32$ new quasars and luminous galaxies at $5.7 < z < 6.8$, of which $24$ were identified as quasars. As the survey already includes quasars with $M_{1450}\sim-22$~mag, extending the quasar luminosity function to this magnitude is now likely within reach. 

\subsection{The VST ATLAS survey}
The VST ATLAS survey\footnote{Webpage VST surveys: https://www.eso.org/public/teles-instr/paranal-observatory/surveytelescopes/vst/surveys/} \citep{Shanks15} is {targeting $5000$~deg$^2$ of the Southern sky \citep{Che18}.} It is pursued through the Very Large Telescope (VLT)\footnote{Webpage VLT: http://www.eso.org/public/teles-instr/paranal-observatory/vlt/}, which consists of four Unit Telescopes with main mirrors of $8.2$~m diameter and four movable auxiliary telescopes. The survey is conducted in the U, V, R, I and Z filters at depths comparable to SDSS. However, it is deeper in the z band, with a mean $5\sigma$ limiting magnitude of $20.89$, and has considerably better seeing, in the range $0.''8-1.''0$ for all five bands.

While the main goal of the survey is to measure baryonic acoustic oscillations, it contributes to the detection of bright $z>6$ quasars. In particular, \citet{Carnall15} reported the discovery of two $z>6$ quasars (selected as i-band dropouts). The first of the quasars in their sample has a redshift $z=6.31\pm0.03$ with $M_{1450}=-27.8\pm0.2$, thus making it the joint second most luminous quasar known at $z>6$. The second quasar corresponds to a redshift of $z=6.02\pm0.03$ with magnitude $M_{1450}=-27.0\pm0.1$. The detection shows the potential of the survey to still discover new very bright quasars at high redshift.

\subsection{The VISTA surveys}
The VISTA surveys\footnote{Webpage VISTA surveys: https://www.eso.org/public/teles-instr/paranal-observatory/surveytelescopes/vista/surveys/}  are conducted via the  Visible and Infrared  Survey Telescope for Astronomy (VISTA)\footnote{Webpage VISTA: http://www.vista.ac.uk/}. VISTA is a 4-m class wide field survey telescope at Paranal in the southern hemisphere, equipped with a near infrared camera (1.65 degree diameter field of view) containing 67 million pixels, available broad band filters at Z,Y,J,H,Ks and a narrow band filter at $1.18$~$\mu$m. For the search for high-redshift quasars, the relevant VISTA surveys are UltraVISTA and VIKING. UltraVISTA is the deepest and narrowest VISTA survey, imaging one patch of sky over and over again to unprecedented depths. The VISTA Kilo-Degree Infrared Galaxy Survey (VIKING) is imaging $1500$~deg$^2$. 

VIKINGS contributed in particular to extend the number of $z>6.4$ quasars, reporting detections at $ z=6.60$, $6.75$, and $6.89$, with black hole masses of $1-2\times10^9$~M$_\odot$ \citep{venemens13}. Within the redshift range $6.44<z<7.44$, they established a lower limit on the black hole mass density of $\rho(M_{BH}>10^9\ M_\odot)>1.1\times10^{-9}$~Mpc$^{-3}$. Jointly with the Kilo-Degree Survey (KiDS)\footnote{Webpage KiDS survey: http://kids.strw.leidenuniv.nl/}, which is pursued by the VLT in the Southern sky, four additional quasars were subsequently discovered at $5.8<z<6.0$, with magnitudes $-26.6 < M_{1450} < -24.4$ \citep{Venemans2015}, thus corresponding to relatively faint objects, and it may be possible to find $30$ objects of similar luminosity during the further course of the survey. Discoveries based on joint data from VISTA and the Dark Energy Survey will further be described below.

\subsection{The Dark Energy Survey}\label{DECam}
The Dark Energy Survey (DES)\footnote{Webpage DES: https://www.darkenergysurvey.org/} consists of a $5000$~deg$^2$ area of the Southern sky (roughly $1/8$ of the total sky), which will be observed over $525$ nights using the new Dark Energy Camera (DECam) mounted on the Blanco~4-meter telescope\footnote{Webpage Blanco telescope: http://www.ctio.noao.edu/noao/node/9} at the Cerro Tololo Inter-American Observatory in the Chilean Andes. While the main science goal is the detection of thousands of supernovae to probe the history of cosmic expansion and thereby the nature of dark energy, the data of the survey have also contributed to the discovery of additional quasars at high redshift.

With DES J0454-4448, the first luminous quasar discovered through the Dark Energy Survey was announced in 2015 \citep{Reed15}, with a redshift of $z = 6.09\pm0.02$ and $M_{1450} = -26.5$, thus corresponding to a rather bright object. Overall, the survey is expected to discover  discover $50-100$ new quasars with $z > 6$ as well as a few with $z > 7$. Indeed, the detection of 8 new quasars, based both on data from DES, VISTA and the Wide-field Infrared Survey Explorer (WISE)\footnote{Webpage WISE: https://www.nasa.gov/mission$\_$pages/WISE/main/index.html}, was reported in 2017 by \citet{Reed17}. This includes the $z = 6.5$ quasar VDES J0224-4711, the second most luminous quasar known with $z \geq 6.5$ ($M_{1450}\gtrsim-27$). The overall redshifts range between $6.0$ and $6.5$, and the magnitude extends to $M_{1450}\lesssim-25$.

\section{Upper limits and constraints from X-ray surveys and observations}\label{xray}
The role and importance of observations in the X-ray regime can hardly be overemphasized to follow the evolution of the supermassive black hole population over cosmic history \citep{Treister10}, as well as the identification of heavily obscured objects that would not be visible at other wavelengths \citep{Treister09}. Surveys pursued with the Chandra satellite\footnote{Webpage Chandra: http://chandra.harvard.edu/}, XMM-Newton\footnote{Webpage XMM-Newton: https://www.cosmos.esa.int/web/xmm-newton} and NuSTAR\footnote{Webpage NuSTAR: https://www.nustar.caltech.edu/} have thus greatly advanced our knowledge on the high-redshift black hole population and demographics, with {newly discovered quasars} up to redshifts of $z\sim5$ \cite{Ueda14}. An excellent review on these discoveries has been provided by \citet{Brandt15}, to which we refer here for a description of the demographics until $z\sim5$. We also emphasize here that the ATHENA X-ray Observatory\footnote{Webpage ATHENA: http://www.the-athena-x-ray-observatory.eu/} has the potential to overcome this boundary, thus potentially providing detections of the first quasars at very high redshift in the future \citep{Nandra13}. {In addition, X-ray observations can be useful to increase available information on already known quasars at $z\gtrsim6$, as demonstrated by \citet{Banados18X}.} In the following, we will discuss how existing X-ray data are already providing strong relevant constraints about the first supermassive black holes.

\subsection{Constraints from the unresolved X-ray background}\label{bg}
The cosmic X-ray background is a radiation background resulting from many individual sources, many of which were initially unresolved, while substantial progress from X-ray surveys has allowed to resolve a substantial fraction of that background. As there is no anticipated cosmological or primordial origin, it is expected that the remaining unresolved background results from low-luminosity sources at low to moderate redshifts, or potentially from higher-luminosity objects at very high redshift, including the population of the very first quasars. With realistic assumptions on the X-ray spectra of the first quasars, the unresolved X-ray background thus translates into a constraint for the first supermassive black holes \citep{Dijkstra04, Salvaterra05}.

To provide constraints on the first quasars, the most relevant component of the unresolved X-ray background is the soft X-ray band at energies of $0.5-2$~keV, which corresponds to hard X-rays at redshifts $z\geq6$. These are well within the regime of energies where the intergalactic medium is optically thin. \citet{Moretti03} have studied the X-ray background in that energy range, finding an integrated energy flux of $\int f_E dE\sim7.53\pm0.35\times10^{-12}$~erg~cm$^{-2}$~s$^{-1}$~deg$^{-2}$ at an energy of $1$~keV, where they combined data from 6 surveys performed by 3 satellites, ROSAT\footnote{Webpage ROSAT: https://heasarc.gsfc.nasa.gov/docs/rosat/rosat3.html}, Chandra, and XMM-Newton. They further included deep pencil beam surveys together with wide field shallow surveys to determine the flux resulting from resolved sources, with individual sources having fluxes of $2.44\times10^{-17}-1.00\times10^{-11}$~erg~cm$^{-2}$~s$^{-1}$. Overall, their analysis could show that $90^{+6}_{-7}\%$ of the soft X-ray background consists of discrete sources.

Point-like sources are however not the only contribution to the X-ray background. Extended emission can originate for instance in clusters and groups of galaxies throughout the Universe, and the expected resulting background was calculated by \citet{Wu01} using the observed X-ray luminosity function, finding an expected contribution of $1.18\times10^{-12}$~erg~cm$^{-2}$~s$^{-1}$, i.e. about $16\%$ of the total soft X-ray background. As argued by \citet{Dijkstra04}, the contribution coming from groups is rather uncertain, so a more conservative estimate may correspond to about $60\%$ of that value, the contribution coming from clusters of galaxies, showing that the current data are not fully consistent. A recent analysis based on the Chandra COSMOS Legacy Survey \citep{Cappelluti17} provides a $1$~keV normalization of the unresolved X-ray background of  $\sim1.37$~keV~cm$^{-2}$~s$^{-1}$~sr$^{-1}$~keV$^{-1}$, stating that unresolved sources contribute $8-9\%$ of the soft background. Even from the unresolved fraction, one would expect that only a small part is really due to the first black holes, thus potentially resulting in a strong upper limit. 

In their analysis, \citet{Dijkstra04} determined that, depending on how to sum up the individual contributions and regarding different scenarios concerning the treatment of the error bars, the unresolved flux is between $0.35-1.23\times10^{-12}$~erg~cm$^{-2}$~s$^{-1}$~deg$^{-2}$, which we will adopt here for definiteness and for comparison with their work. It is however clear that the resulting constraints can be rescaled for different values of the unresolved X-ray background. An additional important ingredient to derive constraints on the first black holes is a quasar spectrum. For this purpose, they adopted the characteristic spectrum derived by \citet{Sazonov04}, who computed the characteristic angular-integrated, broad-band spectral energy distribution for average quasars. This spectrum scales approximately as $f_E\propto E^{-1.7}$ for $E>13.6$~eV, but becomes much shallower beyond $2$~keV, where it scales as $f_E\propto E^{-0.25}$. Through their analysis, \citet{Dijkstra04} have shown that the contribution of quasars to the epoch of reionization is strongly constrained via the unresolved X-ray background, which limits the production rate of ionizing photons. As a result, it became clear that quasars are not the main sources driving reionization, unless their spectra were considerably different at such early times. Partial contributions however still seemed possible, with fractions up to $50\%$ in more optimistic scenarios. Using a similar approach, \citet{Salvaterra05} derived constraints on the mass density of intermediate mass black holes at $z\geq6$, finding an upper limit of $\rho_{BH}<3.8\times10^4$~M$_\odot$~Mpc$^{-3}$, corresponding to about one intermediate mass black hole with 1~Mpc$^{-3}$.

\subsection{Constraints via stacking techniques}
While X-ray observatories have not yet provided any {new discoveries} of supermassive black holes at $z>6$ {(though already known quasars at redshifts $z\gtrsim6$ have been detected with Chandra or XMM, see \citet{Banados18X})}, the results of their surveys can nevertheless be employed to place upper limits on supermassive black holes in typical galaxies. In particular, we recall that the detections in optical surveys are mostly in the higher-luminosity tail of the quasar luminosity function at $z\sim6$. As a result, we obtain statistical information on the properties of the brightest and most massive objects. While these may be interesting as extreme cases to probe potential formation scenarios, it is also important to understand the more typical outcome of black hole and galaxy formation, and to derive constraints on the black hole masses in typical galaxies.

To accomplish this, a stacking techniques has been developed by \citet{Treister11} to combine X-ray data at positions where star-forming galaxies have been found via optical surveys, thus efficiently increasing the signal-to-noise ratio by adding up the data at the different positions. While this of course will not provide information on individual galaxies, it can provide information on the average properties of such star-forming galaxies, either in the form of the detection of a mean signal, or through an upper limit that constrains their X-ray activity. The technique is however technically complex, and was thus subsequently improved by \citet{Willott11} and \citet{Cowie12}. 

Using this technique, \citet{Treister13} employed the $4$~Ms Chandra observations of the Chandra Deep Field-South (CDF-S) \citep{Xue11}, the deepest X-ray observations taken so far, and combined them with additional survey data with galaxy detections at other wavelengths. Within the CDF-S, \citet{Bouwens11} detected $66$ $z\sim7$ galaxies and $47$ at $z\sim8$ through the Hubble Space Telescope (HST)\footnote{Webpage HST: https://www.nasa.gov/mission$\_$pages/hubble/main/index.html} using the WFC3 camera. In addition, HST/WFC3 observations by \citet{Finkelstein12} in the CANDELS fields\footnote{Webpage CANDELS: http://candels.ucolick.org/} yielded a sample of 223 galaxies at $z\sim6$, 80 at $z\sim7$ and 33 at $z\sim8$. While none of these galaxies are detected individually in X-rays, count rates for their positions are nevertheless available from the $4$~Ms Chandra observations by \citet{Xue11}, which can be stacked at the respective positions to increase the signal-to-noise ratio. Using this procedure, \citet{Treister13} obtained an upper limit of the X-ray luminosity of $2.6\times10^{41}$~erg~s$^{-1}$ for the $z\sim6$ sources in the soft band, and $1.6\times10^{42}$~erg~s$^{-1}$ in the hard band. For $z\sim7$ galaxies, these limits correspond to $6.8\times10^{41}$~erg~s$^{-1}$ and $5.3\times10^{42}$~erg~s$^{-1}$, as well as $1.5\times10^{42}$~erg~s$^{-1}$ and $9.8\times10^{42}$~erg~s$^{-1}$ for $z\sim8$ galaxies. All of these are below the standard threshold for active galaxies of $\sim10^{42}$~erg~s$^{-1}$ \citep{Szokoly04}, implying relatively low activity on average.

\begin{figure}
\begin{center}
\includegraphics[width=0.8\textwidth]{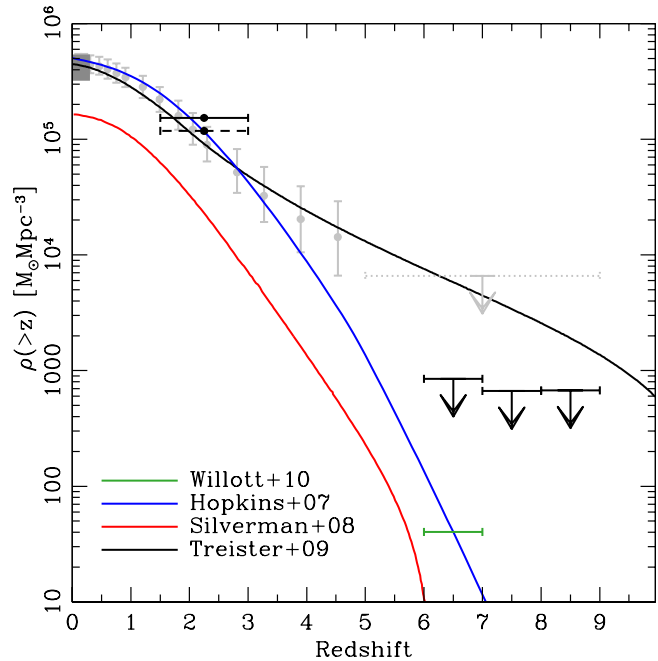}
\end{center}
\caption{Comparison between the observed accreted mass density in supermassive black holes and expectations from the observed luminosity functions, combined with the upper limits from stacking (black solid) and the unresolved soft X-ray background (grey dashed). Adopted from \citet{Treister13},  \textcopyright AAS. Reproduced with permission.}
\label{bhmass_w_z_lf}
\end{figure}

Upper limits on the X-ray luminosity function can be translated into upper limits on the accreted black hole mass density using Soltan's argument \citep{Soltan82}, considering the link between mass accretion and radiation energy production in the accretion process. Applying this analysis to the results from the stacking technique implies upper limits of $990$, $1142$ and $1263$~M$_\odot$~Mpc$^{-3}$ at $z\sim6$, $7$ and $8$, respectively. These limits assume a radiative efficiency of $\epsilon=10\%$, and are even tighter than the ones derived via the unresolved soft X-ray background in section~\ref{bg}, due to the combination of information from a large number of sources.

Translating this into properties of individual galaxies requires further assumptions on the bolometric correction, the Eddington ratio as well as the fraction of active black holes. Assuming $100\%$ activity, a canonical bolometric correction of $10\%$ for hard X-rays and a $10\%$ Eddington ratio, \citet{Treister13} derived an upper limit of the black hole mass of about $3\times10^6$~M$_\odot$ for a typical $z\sim6$ galaxy. This can be higher if not all of them are active, or if the Eddington ratio is further reduced. While the constraints are tighter than originally anticipated, recent research suggests that the limits may still be consistent with total stellar mass - black hole mass relations obtained through calibrations within the local Universe \citep{Volonteri16}. It is nevertheless important to emphasize that black hole formation models need not only reproduce the most luminous quasars discovered via optical surveys, but also have to comply the constraints on the average population that is available from X-ray data.

While the current data do not yet strongly constrain models, the upper limits are important to prevent an overproduction of black holes in certain scenarios. In addition, the main challenge may consist in explaining the most massive quasars that have been observed, through massive seeds, strong subsequent accretion or both. In the two following chapters, we will now address the future prospects, both through the prospects of gravitational wave observatories as well as future observational prospects.

\section*{Acknowledgement}
DRGS thanks for funding through the Anillo Program "Formation and Growth of Supermassive Black Holes" (CONICYT PIA ACT172033). 
{
\bibliographystyle{ws-rv-har}    

}

\printindex[aindx]           
\printindex                  

\end{document}